\documentclass[twocolumn,showpacs,preprintnumbers,amsfonts,amssymb,amsmath]{revtex4}

\usepackage[dvips]{graphicx}
\usepackage{mathrsfs}

\usepackage{graphicx}
\usepackage{hhline}
\usepackage{dcolumn}
\usepackage{bm}

\usepackage[dvips]{color}

\newcommand{\blue}[1]{}

\newcommand{\eref}[1]{Eq.(\ref{#1})}

\newcommand{\psiL}{{\psi_\text{L}}}

\begin{document}


\title{Generating-function approach for bond percolations in hierarchical networks}
\date{\today}

\author{Takehisa Hasegawa}
\email{hasegawa@stat.t.u-tokyo.ac.jp}
\affiliation{Graduate School of Information Science and Technology, The University of Tokyo,
7-3-1, Hongo, Bunkyo-ku, Tokyo, JAPAN.}
\author{Masataka Sato}
\email{hijiri@statphys.sci.hokudai.ac.jp}
\author{Koji Nemoto}
\email{nemoto@statphys.sci.hokudai.ac.jp}
\affiliation{Department of Physics, Graduate School of Science,
Hokkaido University, Kita 10-jo Nisi 8-tyome, Sapporo, JAPAN.}

\begin{abstract}
We study bond percolations on hierarchical scale-free networks with 
the open bond probability of the shortcuts $\tilde{p}$ and that of the ordinary bonds $p$.
The system has a critical phase
in which the percolating probability $P$ takes an intermediate value $0<P<1$.
Using generating function approach,
we calculate the fractal exponent $\psi$ of the root clusters
to show that $\psi$ varies continuously with $\tilde{p}$ in the critical phase.
We confirm numerically that 
the distribution $n_s$ of cluster size $s$ 
in the critical phase obeys a power law $n_s \propto s^{-\tau}$, 
where $\tau$ satisfies the scaling relation $\tau=1+\psi^{-1}$.
In addition the critical exponent $\beta(\tilde{p})$ of the order parameter
varies as $\tilde{p}$, from $\beta\simeq 0.164694$ at $\tilde{p}=0$ to infinity at 
$\tilde{p}=\tilde{p}_c=5/32$.
\end{abstract}


\maketitle

\section{Introduction}

Dynamics on and of complex networks have been one of the focuses of attentions since late 1990s \cite{BarabasiRev,NewmanRev,BoccalettiRev}.
Real networks, e.g., WWW, Internet, food-web, often have complex properties 
such as scale-free degree distribution \cite{SFnet}, 
small-world property \cite{SWnet}, etc., 
to demand more extensive framework of statistical physics 
to investigate the interplay between dynamics and such network topology.
Many analytical and numerical works about the effects of network topology 
on processes such as percolation, interacting spin systems, epidemic processes,
have been reported \cite{DorogoRev}.

Among these issues, unusual phase transitions of percolations and spin systems 
on some networks have attracted our current interests \cite{Callaway,Dorogo01,Kiss03,Coulomb,HN5,Bauer,Kajeh,Hinc,HN4,Berker}.
For example, the percolations on some growing network models undergo 
an infinite order transition with a \textit{Berezinskii-Kosterlitz-Thouless (BKT)-like singularity}: 
(i) the relative size of the largest component vanishes in an essentially singular
way at the transition point, so that the transition is of infinite order, 
and (ii) the mean number $n_s$ of clusters with size $s$ per node
 (or the cluster size distribution in short)
decays in a power-law fashion with $s$,
\begin{equation}
n_s \propto s^{-\tau}, \label{nspower}
\end{equation}
 in a finite region \textit{below} the transition point
 where no giant component exists \cite{Callaway,Dorogo01,Kiss03,Coulomb,HN5}.
A similar non-ordered phase with some power-law behavior, \textit{a critical phase}, 
has also been observed in bond percolations on the enhanced binary tree 
\cite{Nogawa, NAG2, Baek},
 which is one of nonamenable graphs (NAGs) \cite{Lyons00, Schonmann01}.
The system on a NAG takes three distinct phases
according to the open bond probability $p$ as follows: 
(i) the non-percolating phase ($0 \le p \le p_{c1}$) in which only finite size clusters exist, 
(ii) the critical phase ($p_{c1} \le p \le p_{c2}$) in which there are infinitely many infinite clusters, 
and (iii) the percolating phase ($p_{c2} \le p \le 1$) in which the system has a unique infinite cluster.
Here \textit{infinite cluster} means a cluster whose mass diverges with system size $N$ 
as $N^\phi$ with $0<\phi \le 1$. 
To profile the critical phase it is useful to calculate the fractal exponent $\psiL$ defined as 
$s_\text{max} \propto N^\psiL$, where
$s_\text{max}$ is the mean size of the largest components in the system with $N$ nodes.
Note that $\psiL$ corresponds to $d_f/d$  for
 percolating clusters having the fractal dimension $d_f$ on $d$-dimensional Euclidean lattices.
Recent paper \cite{Nogawa} has shown numerically that 
the above phases are characterized as
(i) $\psiL(p) = 0$ for $p<p_{c1}$, 
(ii) continuously increasing of $\psiL(p)$ ($0<\psiL(p)<1$) with $p$, 
where $n_s$ also behaves as (\ref{nspower}) 
with $p$-dependent $\tau$ satisfying
\begin{equation}
\tau =1+\psiL^{-1}, \label{relation}
\end{equation}
for $p_{c1} <p<p_{c2}$, 
and (iii) $\psiL(p)= 1$ for $p>p_{c2}$.
The scaling relation (\ref{relation}) indicates that $\psiL$ plays a role of
the natural cut-off exponent of $n_s$ as shown in the growing random tree
in which $p_{c1}=0$ and $p_{c2}=1$ \cite{HN5}.
In general the growing random networks are considered to have $p_{c1}=0$ 
with finite $p_{c2}$ \cite{Callaway,Dorogo01,Kiss03,Coulomb}.

There exist other systems having a similar phase.
They are in a special class of hierarchical scale-free networks, 
called (decorated) $(u,v)$-flower
introduced comprehensively in \cite{Rozen,RozenNet}.
Berker et al. \cite{Berker} 
have studied bond percolations on the decorated (2,2)-flower by renormalization group (RG)
to show the existence of a \textit{critical} phase 
(as known as the partially ordered phase \cite{Boettcher}), 
where RG flow converges onto the line of nontrivial stable fixed points.
But we have little knowledge about physical properties of the critical phase, i.e., how it is critical.

In this paper, 
we investigate bond percolations on the decorated $(2,2)$-flower with two different probabilities $\tilde{p}$ and $p$, 
which are the open bond probability of the shortcuts and that of the ordinary bonds, respectively.
Here we adopt a generating function approach to calculate 
the fractal exponent, the cluster size distribution, and the order parameter 
for an arbitrary combination of $p$ and $\tilde{p}$,
to reveal a complete picture about the phases of this model.
Our calculations show
(i) the fractal exponent $\psi$ and $\beta$ of the order parameter 
depend on the existing probability $\tilde{p}$ of the shortcuts, 
and (ii) $n_s$ is a power-law at all the point in the critical phase, and its exponent $\tau$ also depends on $\tilde{p}$.

The organization of this paper is as follows:
In Sec.\ref{sec-model},
we introduce the (2,2)-flower and the decorated (2,2)-flower, 
and briefly review the previous studies for the percolation on the flowers \cite{RozenNet, Rozen, Berker}.
In Sec.\ref{sec-method}, 
we introduce the generating functions, 
and derive those recursion relations to calculate the order parameter, fractal exponent, and cluster size distribution. 
The main results are presented in Sec.\ref{sec-result}, 
and Section \ref{sec-summary} is devoted to summary.


\begin{figure}
 \begin{center}
  \includegraphics[width=75mm]{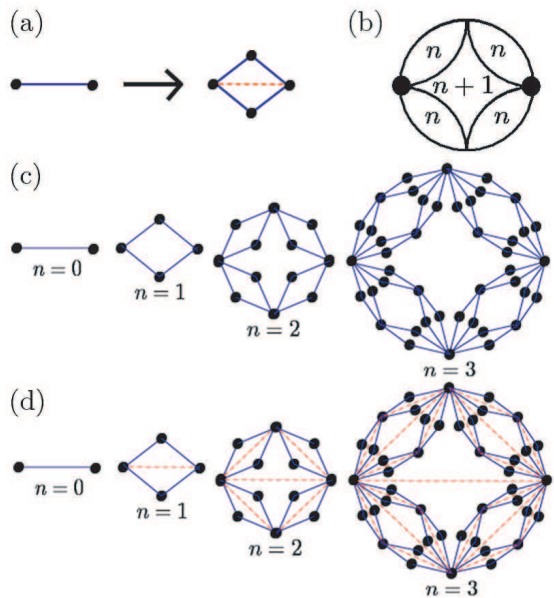}
 \end{center}
 \caption{(Color online)
Construction of the $(2,2)$-flower $F_n$ and the decorated (2,2)-flower $\tilde F_n$.
(a) Each bond is replaced by two parallel paths consisting of two bonds each at next generation. 
(b) The flower $F_{n+1}$ of the  $n+1$-th generation is obtained by joining four copies of $F_n$. 
(c) Realization of $F_n$ with $n=0,1,2,3$.
(d) Realization of $\tilde{F}_n$ with $n=0,1,2,3$. 
$\tilde{F}_n$ is obtained by adding the shortcuts (orange-dashed line in bond replacement (a)) to $F_n$. 
The shortcuts are not replaced by others in each iteration.
}
 \label{rule}
\end{figure}

\begin{figure}
 \begin{center}
  \includegraphics[width=75mm]{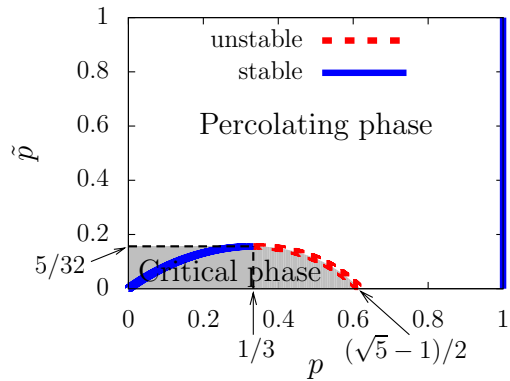}
 \end{center}
 \caption{(Color online)
Phase diagram of the bond percolation on the decorated (2,2)-flower \cite{Berker}.
The blue-solid and the red-dashed lines denote 
$p^*(\tilde p)$ (stable fixed point) and $p_c(\tilde p)$ (unstable fixed point), respectively.
The shaded region represents the critical phase.
}
 \label{phase_diagram}
\end{figure}


\section{Model \label{sec-model}}

In this section we briefly introduce 
the (2,2)-flower and the decorated (2,2)-flower.
The (2,2)-flower $F_n$ of the $n$-th generation is 
constructed recursively as illustrated in Fig.\ref{rule}. 
At $n=0$, the flower $F_0$ consists of two nodes connected by a bond. 
Hereafter we call these nodes \textit{roots}.
For $n \ge 1$, $F_n$ is obtained from $F_{n-1}$, 
such that each existing bond in $F_{n-1}$ is replaced by two parallel paths consisting of two bonds each. 
The decorated (2,2)-flower $\tilde{F}_n$, a variant of the (2,2)-flower $F_n$,
is given by adding shortcuts to $F_n$,
as illustrated in Fig.\ref{rule}(d).

The network properties of these two flowers have been reported in \cite{Rozen,RozenNet}.
The number of nodes $N_n$ of the $n$-th generation is $N_n=2(4^n+2)/3$
 and the degree distribution has 
 a scale-free form $P(k) \propto k^{-\gamma}$ with $\gamma=3$ for both flowers.
One of the important differences between $F_n$ and $\tilde{F}_n$
appears in those dimensionality.
Since the diameter $L_n$ of $F_n$ is $L_n=2^n$, 
the dimension $d$ of the underlying network defined as $N_n \propto L_n^d$ is $2$. 
On the other hand $\tilde{F}_n$ is known to have small-world property $L_n \sim \ln N_n$ 
corresponding to $d \sim \infty$.
In addition $\tilde F_n$ has a high clustering coefficient $C \sim 0.820$, 
in contrast to $C= 0$ for $F_n$.

In the present work we consider the bond percolation on $\tilde F_n$ 
with the open bond probability $p$ of the bonds constituting $F_n$ (the ordinary bonds)
 and that of the shortcuts $\tilde p$ being given independently. 
The standard bond percolation on $F_n$ and $\tilde F_n$ are
given by setting $\tilde{p}=0$ and $\tilde{p}=p$, respectively.
Note that the latter is also given by setting $p=0$ because
$\tilde F_n$ with $p=\tilde p$ and $\tilde F_{n+1}$ with $p=0$ is exactly the same.

The phase diagram is solved exactly by RG technique \cite{Rozen,Berker}.
Let $P^{(n)}$ be a probability that both roots are in the same cluster of a bond configuration on $\tilde{F}_n$ with fixed $\tilde p$.
The initial value is set to $P^{(0)}=p$.
In the large size limit, the system is regarded as percolating if 
$P:=\lim_{n\to\infty} P^{(n)}$ is nonzero.
In this sense $P$ or $P^{(n)}$ is called the \textit{percolation} probability.
Since $P^{(n)}$ is given recursively as 
\begin{eqnarray}
 P^{(n+1)}=1-(1-\tilde{p})\left(1-(P^{(n)})^2\right)^2,  \label{P^{(n)}+1}
\end{eqnarray}
one obtains the flow diagram from the solution of the equation (Fig.\ref{phase_diagram}).
For $0<\tilde{p}<\tilde{p}_c=5/32$, there are two nonzero stable fixed points, $P=p^*(\tilde p)<1$ and $P=1$,
corresponding to the partially ordered phase and the ordered phase, respectively,
 and one unstable fixed point between the two giving the phase boundary, $P=p_c(\tilde p)$.
For $\tilde p >  \tilde p_c$, on the other hand, there is only one stable fixed point at $P=1$,
 so that the system is always percolating.
 
Two special cases, $\tilde{p}=0$ and $\tilde{p}=p$, were investigated in \cite{Rozen}.
Here let us recall their results briefly.
For the case of $\tilde{p}=0$, i.e., 
the standard bond percolation on $F_n$, 
Eq.(\ref{P^{(n)}+1}) 
gives the critical point $p_c(\tilde p=0)=(\sqrt{5}-1)/2$. 
A simple RG argument then gives the critical exponents at $p_c$;
the exponent $\beta\simeq 0.164694$ of the order parameter, i.e.,
the fraction of the largest component $P_\infty \sim (p-p_c)^\beta$, 
and  $\nu\simeq 1.63528$ of the correlation length $\xi \sim |p-p_c|^{-\nu}$,  
which are close to those
 of the two dimensional regular systems. 
On the other hand the same argument for $\tilde{p}=p$ gives 
the infinite order transition, i.e., $\beta\rightarrow\infty$. 

According to the above definition of percolation the percolating cluster should include
both of the root nodes.
It is then convenient to consider the mean size $\langle s_0\rangle_n$
 of the cluster including  both roots (referred to as the root cluster) on $\tilde{F}_n$ instead
of $s_\text{max}(N_n)$ to characterize the criticality:
\begin{equation}
\langle s_0\rangle_n \propto N_n^{\psi},
\end{equation}
where $\psi$ is the fractal exponent for the root cluster.
Note that $\psi$ behaves essentially the same as $\psiL$ for the growing random trees
\cite{HN5}.

\begin{figure}
 \begin{center}
  \includegraphics[width=75mm]{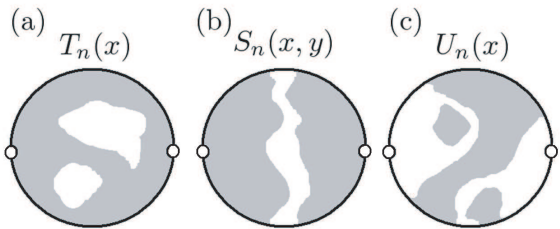}
 \end{center}
 \caption{
Schematic for generating functions
(a) $T_n(x)$, (b) $S_n(x,y)$, and (c) $U_n(x)$.
The open circles represent the root nodes.
}
 \label{description}
\end{figure}
\begin{figure}
 \begin{center}
\includegraphics[width=75mm]{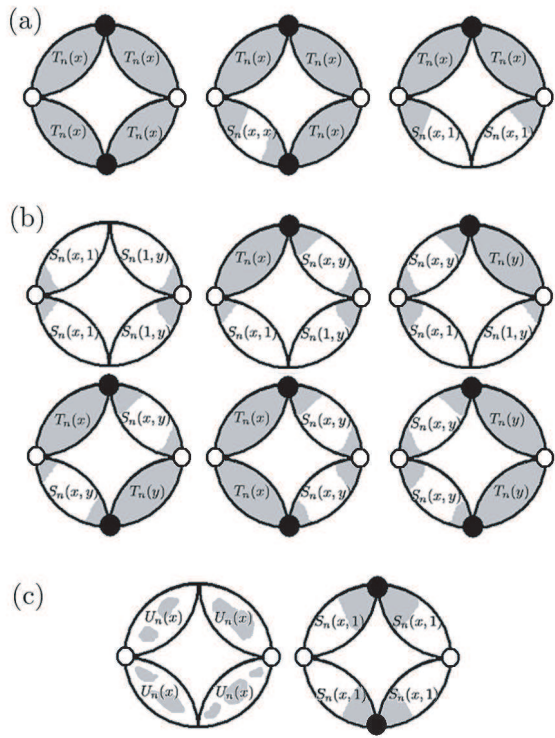}
 \end{center}
 \caption{
Possible diagrams contributing to (a) $T_{n+1}(x)$, (b) $S_{n+1}(x,y)$, (c) $U_{n+1}(x)$. 
The roots (open circles) are not counted in the generating functions,
so that nodes connecting two $F_n$s (closed circles) are taken into account by multiplying 
 $x$ or $y$.
For example the first diagram of (a) represents $x^2T^4_n(x)$ and the fourth diagram of (b)
$xyT_n(x)T_n(y)S^2_n(x,y)$. 
}
 \label{figGenFunc}
\end{figure}


\section{Generating functions \label{sec-method}}

In this section we describe how to utilize generating functions 
to calculate the fractal exponent $\psi$ 
and the cluster size distribution $n_s(p)$ with $\tilde p$ fixed.
First let us consider the bond percolation on $F_n$ with open bond probability $p$ with $\tilde p=0$.
We introduce three basic quantities  on $F_n$:
the probability $t_k^{(n)}(p)$ that both roots are connected
 to the same cluster of size $k$,
the probability $s_{k,l}^{(n)}(p)$ that the left (right) root is connected to a cluster of size
$k$ ($l$) but  these clusters are not the same,
 and the mean number $u_k^{(n)}(p)$ of clusters of size $k$ to which neither of the roots
 is connected.
For the sake of convenience the roots are not counted in the cluster size $k$ or $l$
 for $t_k^{(n)}(p)$ and $s_{k,l}^{(n)}(p)$.
The corresponding generating functions are defined as
\begin{subequations}
\label{TSU_n}
\begin{eqnarray}
 T_n(x)&=&\sum_{k=0}^{\infty}t_k^{(n)}(p)x^k,  \label{T_n}\\
 S_n(x,y)&=&\sum_{k=0}^{\infty}\sum_{l=0}^{\infty}s_{k,l}^{(n)}(p)x^ky^l,
  \label{S_n}\\
 U_n(x)&=&\sum_{k=0}^{\infty}u_k^{(n)}(p)x^k.  \label{U_n}
\end{eqnarray}
\end{subequations}
The self-similar structure of  $F_n$ allows us to obtain 
the recursion relations for the above generating functions 
\begin{subequations}
\label{TSU_n+1_old}
\begin{eqnarray}
 T_{n+1}(x)&=&x^2T_n^4(x)+4x^2T_n^3(x)S_n(x,x) \nonumber \\
&{}&+2xT_n^2(x)S_n^2(x,1),
  \label{T_n+1_old} \\
 S_{n+1}(x,y)&=&S_n^2(x,1)S_n^2(1,y) \nonumber \\
&{}&+2S_n(x,y)S_n(x,1)S_n(1,y) \nonumber \\
&{}&\phantom{+}\times \left[xT_n(x)+yT_n(y)\right] \nonumber \\
&{}&+2xyS_n^2(x,y)T_n(x)T_n(y) \nonumber \\
&{}&+S_n^2(x,y)\left[x^2T_n^2(x)+y^2T_n^2(y)\right],
  \label{S_n+1_old} \\
 U_{n+1}(x)&=&4U_n(x)+2xS_n^2(x,1),
  \label{U_n+1_old}
\end{eqnarray}
\end{subequations}
as illustrated in Fig.\ref{description} and Fig.\ref{figGenFunc}. 
The initial conditions are given as $T_0(x)=p$, $S_0(x,y)=q=1-p$ and $U_0(x)=0$.

It is convenient to rewrite these recursion formulas in terms of functions of single variable $x$ only.
In order to this we introduce new functions
$V_n(x)\equiv S_n(x,x)$ and $R_n(x)\equiv S_n(x,1)$ to obtain
\begin{subequations}
\label{TVRU_n+1}
\begin{eqnarray}
T_{n+1}(x)
&=&\mathscr{T}\left[T_n(x),V_n(x),R_n(x),x\right] \nonumber \\
&\equiv&x^2T_n^4(x)+4x^2T_n^3(x)V_n(x) \nonumber \\
&{}&+2xT_n^2(x)R_n^2(x),   \label{T_n+1} \\
V_{n+1}(x)
&=&\mathscr{V}\left[T_n(x),V_n(x),R_n(x),x\right] \nonumber \\
&\equiv&R_n^4(x)+4xT_n(x)V_n(x)R_n^2(x) \nonumber \\
&{}&+4x^2T_n^2(x)V_n^2(x),  \label{V_n+1} \\
R_{n+1}(x)
&=&\mathscr{R}\left[T_n(x),R_n(x),x\right] \nonumber \\
&=&R_n^2(x)\left[1+xT_n(x)\right]^2,  \label{R_n+1} \\
U_{n+1}(x)
&=&\mathscr{U}\left[R_n(x),U_n(x),x\right] \nonumber \\
&\equiv&4U_n(x)+2xR_n^2(x).  \label{U_n+1}
\end{eqnarray}
\end{subequations}
Indeed it is this form that enables us to obtain the solutions for large $n$ numerically.

Now the construction of the recursion relations for $\tilde F_n$ 
with $\tilde{p}$ fixed is straightforward.
Let $\tilde{T}_n(x)$, $\tilde{V}_n(x)$, $\tilde{R}_n(x)$ and $\tilde{U}_n(x)$
denote the corresponding generating functions on $\tilde F_n$.
By using the above formula (\ref{T_n+1})-(\ref{U_n+1}) with these functions one can construct the generating functions
$T_{n+1}(x)$, $V_{n+1}(x)$, $R_{n+1}(x)$ and $U_{n+1}(x)$
 on the decorated (2,2)-flower of the next generation \textit{without}
 the shortcut directly connecting the roots:
\begin{subequations}
\label{TVRU_itar}
\begin{eqnarray}
 T_{n+1}(x)&=&\mathscr{T}\left[\tilde{T}_n(x),\tilde{V}_n(x),\tilde{R}_n(x),x\right],
  \label{T_itar}\\
 V_{n+1}(x)&=&\mathscr{V}\left[\tilde{T}_n(x),\tilde{V}_n(x),\tilde{R}_n(x),x\right],
  \label{V_itar}\\
 R_{n+1}(x)&=&\mathscr{R}\left[\tilde{T}_n(x),\tilde{R}_n(x),x\right],
  \label{R_itar}\\
 U_{n+1}(x)&=&\mathscr{U}\left[\tilde{R}_n(x),\tilde{U}_n(x),x\right]. 
  \label{U_itar}
\end{eqnarray}
\end{subequations}
The flower $\tilde F_{n+1}$ is made by adding the shortcut to the intermediate one with
probability $\tilde{p}$ and one thus obtains 
\begin{subequations}
\label{TVRU_dec}
\begin{eqnarray}
 \tilde{T}_{n+1}(x)&=&T_{n+1}(x)+\tilde{p}V_{n+1}(x),
  \label{T_dec}\\
 \tilde{V}_{n+1}(x)&=&\tilde{q}V_{n+1}(x),
  \label{V_dec} \\
 \tilde{R}_{n+1}(x)&=&\tilde{q}R_{n+1}(x),
  \label{R_dec}\\
 \tilde{U}_{n+1}(x)&=&U_{n+1}(x),
  \label{U_dec}
\end{eqnarray}
\end{subequations}
where $\tilde{q}=1-\tilde{p}$.
The initial conditions are given as 
$\tilde{T}_0(x)=p$, $\tilde{V}_{0}(x)=\tilde{R}_{0}(x)=q$ and $\tilde{U}_0(x)=0$.
One can easily check the probability conservation $\tilde T_n(1)+\tilde V_n(1) = 1$
 by the iteration (\ref{TVRU_dec}).

Once these generating functions are obtained one can evaluate various quantities of the present interest.
For example the percolation probability $P^{(n)}$ is given as
\begin{subequations}\label{Pn_Qn}
\begin{align}
 P^{(n)}&=\tilde T_n(1),\\
 Q^{(n)}&\equiv 1-P^{(n)}=\tilde S_n(1,1)=\tilde V_n(1)=\tilde R_n(1). 
\end{align}
\end{subequations}
Note that (\ref{P^{(n)}+1}) is re-obtained
 by putting $x=1$ to (\ref{TVRU_dec}) and using (\ref{Pn_Qn}).
The mean number of the root cluster $\langle s_0 \rangle_n$  (or the
order parameter $P_\infty^{(n)}(p)$)
and the cluster size distribution $n_s^{(n)}(p)$ 
on $\tilde F_n$  are given as
\begin{eqnarray}
 \langle s_0 \rangle_n
&=&\tilde T_n'(1)+\tilde V_n'(1),  \label{s_0}\\
 P_{\infty}^{(n)}(p)&=&\frac{\langle s_0 \rangle_n}{N_n} 
= \tau_n + \sigma_n, \label{P_inf}\\
 n_s^{(n)}(p)&=&\frac{\tilde u_s^{(n)}(p)}{N_n},  \label{n_s}
\end{eqnarray}
where the prime denotes the first derivative with respect to $x$,
and we put $\tau_n=\tilde T_n^{\prime}(1)/N_n$ and $\sigma_n=\tilde V_n^{\prime}(1)/N_n$.

It is useful to 
consider the derivatives of recursion relations (\ref{T_dec})-(\ref{R_dec})
 for evaluating $P_\infty^{(n)}(p)$.
By noticing
 $\tilde V_n^{\prime}(1)=2\tilde R_n^{\prime}(1)$
we obtain the recursion relations for $\tau_n$ and $\sigma_n$ as
\begin{widetext}
 \begin{eqnarray}
\begin{pmatrix}
    \sigma_{n+1} \\
    \tau_{n+1} 
\end{pmatrix}
  &=&
  \frac{N_n}{N_{n+1}}
\begin{pmatrix}
  2\tilde{q}Q^{(n)}(1+P^{(n)})^2  & 4\tilde{q}(Q^{(n)})^2(1+P^{(n)}) \\
  2(1+P^{(n)})\left[(P^{(n)})^2+\tilde{p}Q^{(n)}(1+P^{(n)})\right] & 4\left[1-\tilde{q}(Q^{(n)})^2(1+P^{(n)})\right]\
\end{pmatrix}
\begin{pmatrix}
    \sigma_{n} \\
    \tau_{n} 
\end{pmatrix}
 \nonumber \\
  &{}&+\frac{1}{N_{n+1}}
\begin{pmatrix}
    4\tilde{q}P^{(n)}(Q^{(n)})^2(1+P^{(n)}) \\
    2P^{(n)}\left[2-P^{(n)}-2\tilde{q}(Q^{(n)})^2(1+P^{(n)})\right] 
\end{pmatrix}
 \label{Matrix}\\
  &\simeq&
\begin{pmatrix}
    \frac{1}{2}\tilde{q}Q(1+P)^2 & \tilde{q}Q^2(1+P) \\
    \frac{1}{2}(1+P)\left[P^2+\tilde{p}Q(1+P)\right] & 1-\tilde{q}Q^2(1+P) 
   \end{pmatrix}  
   \begin{pmatrix}
    \sigma_{n} \\
    \tau_{n} 
   \end{pmatrix}  
  \quad \mbox{(for $n \gg 1$)},
  \label{Matrix_inf}
 \end{eqnarray}
\end{widetext}
where we recall
$P=\lim_{n\rightarrow\infty}P^{(n)}$ and
 $Q=\lim_{n\rightarrow\infty}Q^{(n)}$.
Note that this expression is an extension of Eq.(31) in \cite{Rozen}.


\section{Results \label{sec-result}}

\begin{figure}
 \begin{center}
\includegraphics[width=75mm]{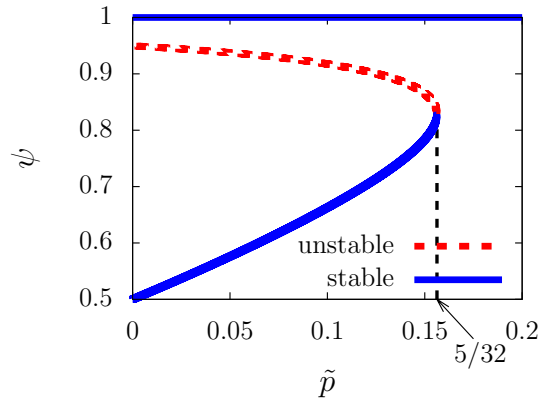}
 \end{center}
 \caption{(Color online)
Fractal exponent $\psi$ on the stable fixed points $p^*(\tilde p)$ (blue-solid line) 
and unstable fixed points $p_c(\tilde p)$ (red-dashed line).
}
 \label{figPsi}
\end{figure}

To profile the critical phase we calculate the fractal exponent $\psi$. 
In the ordered phase we have trivially $\psi=1$. Otherwise
the fixed points $P (<1)$ of the RG equation (\ref{P^{(n)}+1}) 
satisfy $\tilde{q}(1-P)(1+P)^2=1$.
The recursion relation (\ref{Matrix_inf}) is then reduced to 
\begin{equation}
   \begin{pmatrix}
    \sigma_{n+1} \\
    \tau_{n+1} 
   \end{pmatrix}  
=
\begin{pmatrix}
   \frac{1}{2} & \alpha\\
   \frac{1}{2}P & 1-\alpha
  \end{pmatrix}  
   \begin{pmatrix}
    \sigma_{n} \\
    \tau_{n} 
   \end{pmatrix},
\end{equation}
where $\alpha=(1-P)/(1+P)$.
By using the largest eigenvalue $\lambda(P)$ of the above matrix,
\begin{eqnarray}
 \lambda(P)=\frac{1}{4}\left[(3-2\alpha)+\sqrt{1-4\alpha(1-2P)+4\alpha^2}\right],
  \label{eigen_value}
\end{eqnarray}
we can calculate the fractal exponent $\psi$ on the fixed points in the same way as \cite{Rozen} does:
\begin{eqnarray}
 \psi(P)=1+\frac{\ln{\lambda(P)}}{\ln{4}}.
  \label{psi}
\end{eqnarray}
The $\tilde{p}$-dependence of $\psi$ is shown in Fig.\ref{figPsi}.
We find that 
(i) for $\tilde{p}<\tilde p_c=5/32$,
 $\psi$ on the (un)stable fixed points increases (decreases) with increasing $\tilde{p}$, 
and (ii) for $\tilde{p}>\tilde p_c$, $\psi$ is equal to one irrespective of both $p$ and $\tilde{p}$, 
which means that the system is always in the percolating phase.
Let us consider the $p$-dependence of $\psi$ with $\tilde p < \tilde p_c$ fixed.
In the critical phase $( p < p_c(\tilde p))$ the percolation probability $P^{(n)}$ goes to $p^*(\tilde p)$ and
 the exponent $\psi=\psi(p^*(\tilde p))$ is constant in this region.
At the critical point $(p=p_c(\tilde p))$ the fixed point is $P=p_c(\tilde p)$ itself
 and thus $\psi$ discontinuously changes to $\psi(p_c(\tilde p))$, and jumps again to one for
the percolating phase $(p>p_c(\tilde p))$.
This behavior can be also confirmed directly by evaluating the $N_n$-dependence of $\langle s_0\rangle_n$
numerically (not shown).
This result indicates that the probability $\tilde{p}$ of the shortcuts essentially determines how the system is  \textit{critical} in the partially ordered critical phase.
On the other hand, for the standard bond percolation on the decorated $(2,2)$-flower ($\tilde{p}=p$), 
the fractal exponent $\psi$ varies continuously with open bond probability $p$ as observed on a NAG \cite{Nogawa}. 

\begin{figure}[!h]
 \begin{center}
  \includegraphics[width=75mm]{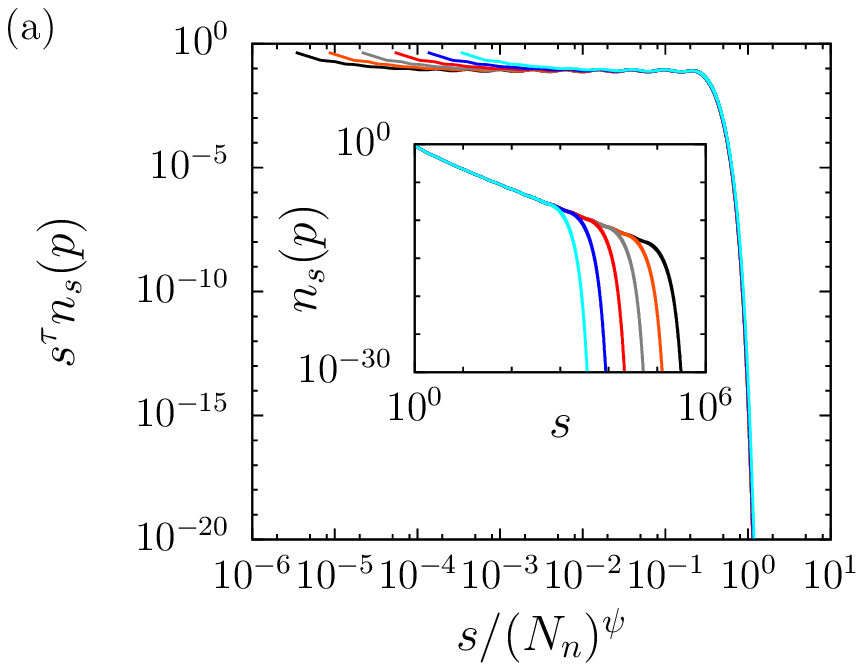}
  \includegraphics[width=75mm]{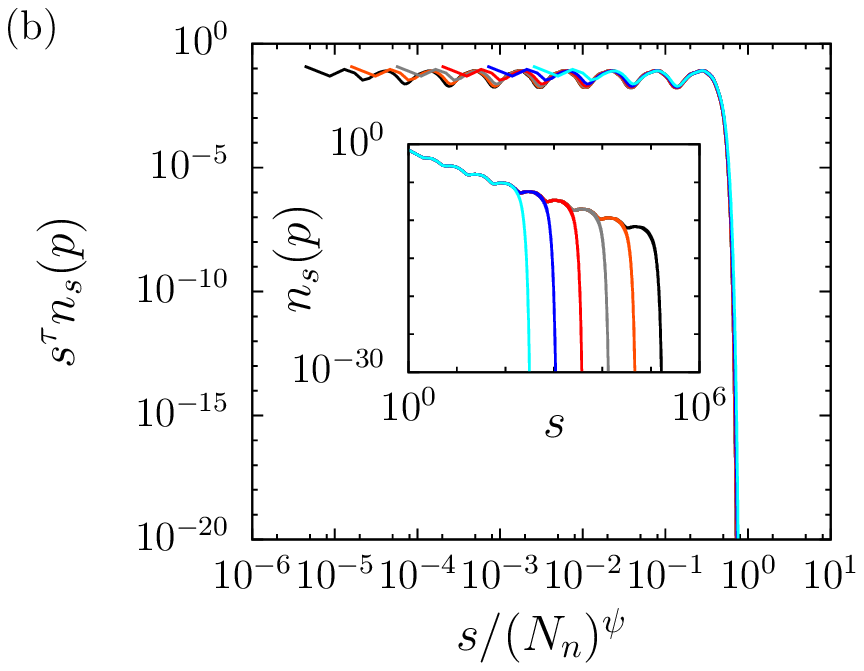}
  \includegraphics[width=75mm]{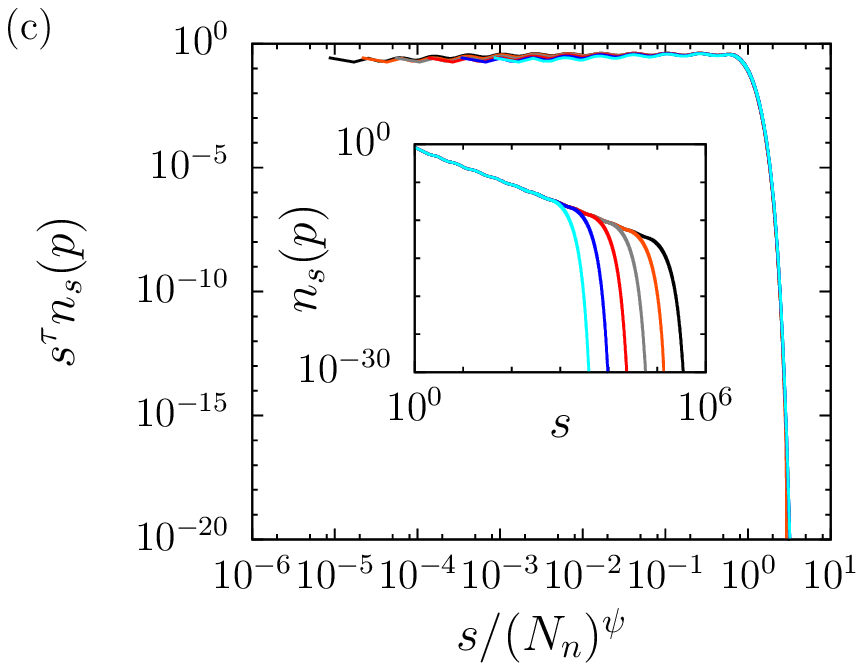}
 \end{center}
 \caption{
(Color online)
Finite size scaling for cluster size distribution $n_s$ at (a) stable fixed point ($p=0.130302$) and (b) unstable fixed point ($p=0.517492$), and (c) $p=0.3$, for $\tilde{p}=0.1$. 
Here $\psi$ and $\tau$ in (c) are given by 
those at the stable fixed point.
Insets show raw data of $n_s$. 
We set (a) $n=9,10,11,12,13,14$, 
(b) $n=5,6,7,8,9,10$, and (c) $n=8,9,10,11,12,13$, from left to right.
}
 \label{figNs}
\end{figure}
\begin{figure}
 \begin{center}
  \includegraphics[width=75mm]{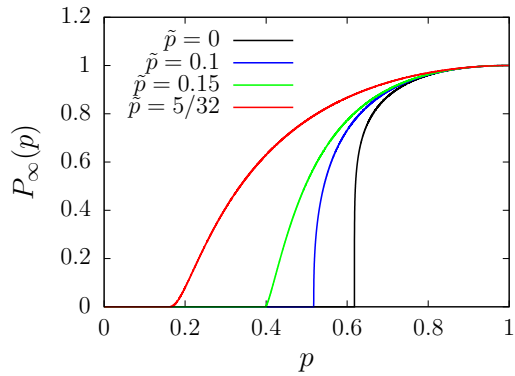}
 \end{center}
 \caption{
(Color online)
Fraction of the largest components $P_{\infty}(p)$ on $\tilde F_n$ 
with $\tilde{p}=0,0.1,0.15,$ and $5/32(=\tilde p_c)$ (from right to left). 
Here $n$ is taken to $10^6 (\gg 1)$.
}
 \label{figP_inf}
\end{figure}
\begin{figure}
 \begin{center}
\includegraphics[width=75mm]{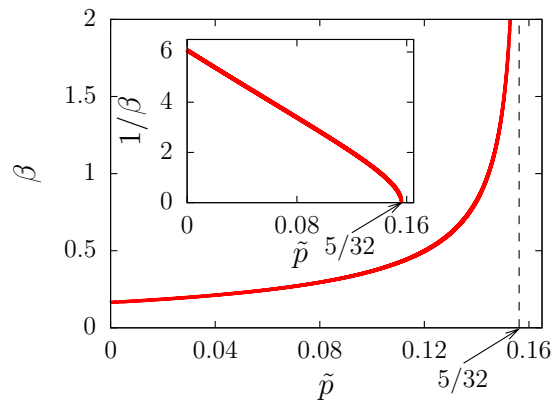}
 \end{center}
 \caption{
$\tilde p$-dependence of the critical exponent $\beta$ on the phase boundary $p=p_c(\tilde p)$.
Inset shows that of $\beta^{-1}$.
}
 \label{figBeta}
\end{figure}

We also evaluate numerically the recursion relations (\ref{T_dec})-(\ref{U_dec}) 
to obtain the cluster size distribution $n_s$ on $\tilde{F}_n$ given by (\ref{n_s}). 
We should observe a power-law behavior of $n_s$ in the whole region of the critical phase.
To check this behavior we assume a finite size scaling form
\begin{eqnarray}
n_s(N) = s^{-\tau} f( s  N^{-\psi} ), 
\end{eqnarray}
where $\psi$ is the fractal exponent obtained above and the scaling function $f(x)$ behaves as
\begin{equation}
f(x)\sim
\begin{cases}
\text{rapidly decaying func.} & \text{for $x\gg1$,}\\
\text{constant} & \text{for $x\ll 1$,}
\end{cases}
\end{equation}
and $\tau$ satisfies the scaling relation \cite{HN5,Nogawa}
\begin{eqnarray}
 \tau=1+\psi^{-1}.
  \label{tau-psi}
\end{eqnarray}
As discussed in \cite{HN5}
a plausible argument leads us to the relation as follows: 
By assuming $n_s \propto s^{-\tau}$ 
 in the critical phase, 
\textit{a natural cutoff} $s_\text{max}$ of the cluster size distribution 
(a natural cutoff of the degree distribution was introduced in \cite{DorogoCutoff}) is given as 
\begin{equation}
N \int_{s_\text{max}}^{\infty} n_s ds \simeq 1 \to s_\text{max} \propto N^{\frac{1}{\tau-1}}.
\end{equation}
Here we emphasize that $s_\text{max}$ plays just a role of characteristic cutoff of the distribution
and does not need to be strictly  the mean size of the largest clusters.
In this sense 
we can replace the fractal exponent $\psi_\text{L}$ of $s_\text{max}$ with $\psi$ of $\langle s_0 \rangle$, so that the relation (\ref{tau-psi}) follows.
Indeed the largest cluster \textit{not} containing the roots is expected to 
the one containing the top-most (or, equivalently, bottom-most) node in the right figure of Fig.\ref{figGenFunc}(c) contributing the second term of (\ref{U_n+1_old}).
The mean size of the cluster is proportional to $\tilde V'_n(1)$ and so to $\langle s_0\rangle$.
Therefore one can conclude that the characteristic size 
grows with $N$ not faster than  $N^\psi$.

Our finite size scaling for $n_s$ is indeed well fitted
 on both stable and unstable fixed points as shown in Fig.\ref{figNs}. 
Note that  for the sake of (\ref{tau-psi}) no fitting parameter remains.
The scaling also works at any $p$ in the critical phase (Fig.\ref{figNs}(c)),
but the convergence is not so rapid as on the fixed points.

Finally we iterate \eref{Matrix_inf} numerically to obtain the order parameter $P_\infty^{(n)}(\tilde{p})$ 
on $\tilde F_n$.
The result for $n=10^6$
is shown in Fig.\ref{figP_inf}.
The initial growth of the order parameter becomes moderate with increasing
$\tilde p$.
To examine the critical exponent $\beta$ on the phase boundary $p=p_c(\tilde p)$, 
we follow the scaling argument in \cite{Rozen} to obtain 
\begin{equation}
 \beta(\tilde{p})=
-\frac{\ln{\lambda(p_c(\tilde{p}))}}{\ln{\Lambda(p_c(\tilde p))}},
  \label{beta}
\end{equation}
where
\begin{equation}
 \Lambda(P)=\left.\frac{\partial P^{(n+1)}}{\partial P^{(n)}}\right|_{P}
  =4\tilde{q}P(1-P^2)=2(1-\alpha).
  \label{Lambda}
\end{equation}
Figure \ref{figBeta} shows the $\tilde{p}$-dependence of $\beta$.
We find that $\beta$ increases continuously with $\tilde{p}$, 
from $\beta =0.164694$ at $\tilde{p}=0$ to $\beta=\infty$ at $\tilde{p}=\tilde p_c=5/32$. 
At $\tilde p=\tilde{p}_c$ we expand \eref{P^{(n)}+1} near $p_c(\tilde p_c)=1/3$ to obtain
\begin{equation}
 \Delta P^{(n+1)}\simeq\Delta P^{(n)}+\frac{9}{8}(\Delta P^{(n)})^2,
 \label{DeltaP}
\end{equation}
where $\Delta P^{(n)}=P^{(n)}-p_c(\tilde p_c)$.
We can estimate the solution for small $\Delta P^{(0)}=p-p_c(\tilde p_c)>0$ as
\begin{equation}
\Delta P^{(n)}\simeq\Delta P^{(0)}+\frac{9n}{8}(\Delta P^{(0)})^2,
\end{equation}
which is correct as long as the second term in the r.h.s. is much less than the first one, or equivalently,
$n$ is much less than $n^*\simeq 1/\Delta P^{(0)}=|p-p_c|^{-1}$.
For $n\gtrsim n^*$, $P^{(n)}$ goes to 1 rapidly and so we obtain
\begin{equation}
P_{\infty}\sim \lambda^{n^*}\sim \exp\left(-\frac{\mbox{const.}}{p-p_c}\right),
\end{equation}
where $\lambda=\lambda(p_c(\tilde p_c))$ given by \eref{eigen_value}.
We thus find an essential singularity in the order parameter at $\tilde{p}=\tilde p_c$.

Note that $\beta$ is apparently related to $\psi$ through $\lambda(p_c(\tilde p))$
as shown in \cite{Rozen} (see Eqs.(\ref{psi}) and (\ref{beta})).
It is, however, not the case in the (off-boundary) critical phase where
the nontrivial stable fixed points $p^*(\tilde p)$ dominate the criticality while
the order parameter vanishes there.


\section{Summary \label{sec-summary}}

We have investigated bond percolations on the decorated (2,2)-flower with two 
different probabilities $p$ and $\tilde{p}$. 
Our generating function approach has revealed that 
the system is in the critical phase for $p<p_c(\tilde{p})$ and $\tilde{p} <\tilde p_c = 5/32$. 
We have evaluated the fractal exponent $\psi$ and 
confirmed the power-law behavior of $n_s$ in the critical phase 
as well as those dependence on $p$ and $\tilde{p}$ and the validity of the scaling relation
$\tau = 1 + \psi^{-1}$.

We have also examined the critical exponent $\beta$ in the percolating phase and found 
that $\beta$ also varies as $\tilde{p}$ 
from $\beta\simeq 0.164694$ at $\tilde{p}=0$, where the network is two-dimensional-like, 
to $\beta = \infty$ at $\tilde{p}=\tilde p_c$,
 where the dimensionality of the underlying network is infinite.

It is only at $\tilde p=\tilde p_c$ that
 the percolation on the decorated (2,2)-flower shows
 an infinite order transition with the BKT-like singularity as percolations on growing networks do \cite{Callaway,Dorogo01,Kiss03,Coulomb}.
The finiteness of $\beta$ for $\tilde p<\tilde p_c$ suggests that
the existence of some critical phase adjacent to the \textit{normal} ordered phase is not enough
 for the network to have such an essential singularity in the order parameter
 and thus an infinite order phase transition.
At present we have none of the key to reveal necessary conditions for the existence
of the BKT-like singularity.
Further study would be required to clarify the relation between these interesting properties of
the phase transition.

\begin{acknowledgments} 
The authors thank Tomoaki Nogawa for helpful discussions.
\end{acknowledgments}

\end{document}